\documentclass[twocolumn,prb,aps,epsfig,showpacs,floatfix,footinbib]{revtex4-1}
\usepackage{graphics}
\usepackage{graphicx}
\usepackage{dcolumn} 
\usepackage{bm}
\usepackage{epsfig}
\usepackage{natbib}
\usepackage{lineno}

\begin{document}

\title{Large quantum fluctuations in the strongly coupled spin-1/2
chains of green dioptase: a hidden message from birds and trees}

\author{O. Janson}
\email{janson@cpfs.mpg.de}
\affiliation{Max-Planck-Institut f\"{u}r Chemische Physik fester
Stoffe, D-01187 Dresden, Germany}

\author{A. A. Tsirlin}
\affiliation{Max-Planck-Institut f\"{u}r Chemische Physik fester
Stoffe, D-01187 Dresden, Germany}

\author{M. Schmitt}
\affiliation{Max-Planck-Institut f\"{u}r Chemische Physik fester
Stoffe, D-01187 Dresden, Germany}


\author{H. Rosner}
\email{rosner@cpfs.mpg.de}
\affiliation{Max-Planck-Institut f\"{u}r Chemische Physik fester
Stoffe, D-01187 Dresden, Germany}

\date{22$^{\text{nd}}$ of April, 2010 --- dedicated to Stefan-Ludwig Drechsler on the occasion
of his 60$^{\text{th}}$ birthday  }
\begin{abstract}
The green mineral dioptase
Cu$_6$Si$_6$O$_{18}$$\cdot$6H$_2$O has been known
since centuries and plays an important role in esoteric doctrines. In
particular, the green dioptase is supposed to grant the skill to speak with trees and
to understand the language of birds. Armed with natural samples of
dioptase, we were able to unravel the magnetic nature of the mineral
(presumably with hidden support from birds and trees)
and show that strong quantum fluctuations can be realized in an essentially
framework-type spin lattice of coupled chains, thus neither
frustration nor low-dimensionality are prerequisites. We present a
microscopic magnetic model for the green dioptase.  Based on
full-potential DFT calculations, we find two relevant couplings in
this system: an antiferromagnetic coupling $J_c$, forming spiral
chains along the hexagonal $c$ axis, and an inter-chain ferromagnetic
coupling $J_d$ within structural Cu$_2$O$_6$ dimers. To refine the
$J_c$ and $J_d$ values and to confirm the proposed spin model, we
perform quantum Monte-Carlo simulations for the dioptase spin
lattice. The derived magnetic susceptibility, the magnetic ground
state, and the sublattice magnetization are in remarkably good agreement
with the experimental data. The refined model parameters are
$J_c$=78~K and $J_d$=$-$37~K with $J_d/J_c\simeq -0.5$. Despite the
apparent three-dimensional features of the spin lattice and the lack
of frustration, strong quantum fluctuations in the system are
evidenced by a broad maximum in the magnetic susceptibility, a reduced
value of the N\'eel temperature $T_N\simeq15$\,K $\ll J_c$, and a low value of the
sublattice magnetization $m$=$0.55$~$\mu_{B}$. All these features should be
ascribed to the low coordination number of 3 that outbalances the
three-dimensional nature of the spin lattice.
\end{abstract}

\maketitle

\section{Introduction}
\begin{figure*}
\begin{center}\includegraphics[width=5cm,angle=270]{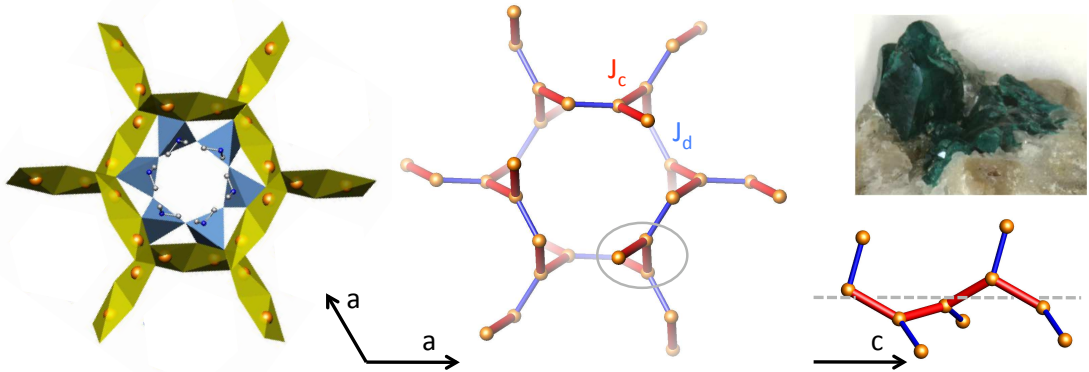}\end{center}
\caption{\label{struc}(Color online)
Left: Crystal structure of the green dioptase. The
Cu$_2$O$_6$ dimers are shown in yellow and form a 3D
network. SiO$_4$ tetrahedra are shown blue. The crystal
water is depicted by O (blue) and H (gray) atoms with O--H bonds.  
Middle: the magnetic model of the green dioptase. Cu atoms
are depicted as orange spheres, other atoms are not shown.
The leading antiferromagnetic coupling $J_c$ (red) forms spiral
chains running along $c$ perpendicular to the projection plane. The
ferromagnetic coupling $J_d$ (blue) within the structural Cu$_2$O$_6$
dimers couples the chains into a three-dimensional
framework. Right: section of the spiral chain along $c$ (bottom) and
a natural sample of green dioptase grown on calcite (top). 
}\end{figure*}

Since ancient times, emerald has been one of the most rare and treasured
gemstones because of its bright and brilliant green color. However, by
far not all gemstones that were collected as emeralds or varieties of
it were indeed emeralds --- many of them later appeared to be specimens
of green dioptase (see Fig.~\ref{struc}). Nevertheless, this confusion
contributed considerably to the assignment of many mysterious powers
to this mineral, such as providing beauty, wealth and
creativity. Especially, esoteric doctrines credit dioptase with the
ability to grant the skill to speak with trees and understand the
language of birds. Stimulated by such a rather complex realm of
concealed powers, we attempt in the present study to unravel the also
controversially debated magnetic properties based on the paradigm of
quantum mechanics and modern electronic structure theory.

Dioptase is a copper silicate mineral forming remarkably large shiny
green rhombohedral crystals. Scientifically, it was first described
and named by Ren\'e-Just Ha\"uy in the famous "Trait\'e de
Min\'eralogie" in 1801.\cite{hauy} Vauquelin, as reported in
Ref.~\onlinecite{hauy}, found that dioptase was a copper mineral
containing silicate and --~erroneously~-- carbonate anions. Only later, pure
dioptase samples were analyzed and recognized as copper silicate
with crystal water.\cite{hess} 

Structure determination showed that hydrous dioptase with the chemical
composition Cu$_6$Si$_6$O$_{18}$$\cdot$6H$_2$O is a cyclosilicate
with 6-membered single rings of silica Si$_6$O$_{18}$ (compare
Fig.~\ref{struc}, left) crystallizing in the space group $R\bar3$
(S. G. 148).\cite{heide,HC_CuSiO3H2O_ENS} These rings are interconnected by
Cu$^{2+}$ ions with a characteristic local environment of elongated
octahedra formed by oxygen atoms. The Cu--O bond length in this 4+2
arrangement is quite typical with about 1.96
\AA\ for the distorted equatorial plane of the octahedra and a bit
larger than usual (between 2.6 and 2.75 \AA ) for the apical oxygens
belonging to the crystal water.  Due to the sharing of the octahedral
O--OH$_2$ edges, the magnetic Cu$^{2+}$ ions form helical chains around
the 3-fold axis along $c$ (see Fig.~\ref{struc}). Thus, each Cu atom
has two Cu neighbors along these chains and another Cu neighbor with
which it forms an edge-shared Cu$_2$O$_6$ dimer that connects two
adjacent spiral chains (see Fig.~\ref{struc}).

The magnetic properties of green dioptase have been investigated in
some experimental
studies.\cite{HC_CuSiO3H2O_spence58,HC_CuSiO3H2O_newnham67,wintenberger1993,
HC_CuSiO3H2O_CpT_TG,HC_CuSiO3H2O_ENS} Although these studies yield
quantitatively slightly varying results,
likely also related to dioptase samples originating from different
locations, they essentially converge in the description of dioptase as
an antiferromagnet with a rather low N\'eel temperature ($T_N \sim$ 15
K) compared to the antiferromagnetic Curie-Weiss temperature of about
70 K. The ordered magnetic moment ($m = 0.55 \mu_B$) is drastically
reduced with respect to the saturation moment of 1 $\mu_B$ for
Cu$^{2+}$. Together with the broad maximum in the measured magnetic
susceptibility,\cite{HC_CuSiO3H2O_ESR_chiT} this puts the compound in the family of spin 1/2 quantum
magnets that can be described successfully in many cases by the
isotropic Heisenberg model

\begin{equation}	
\hat{H}=\sum_{<ij>}J_{ij}\hat{S_i}\hat{S_j},
\end{equation}

at least for the low lying spin excitations. Here, $J_{ij}$ represents
the exchange interaction between spins located at the lattice sites
$i$ and $j$.

Although this model looks deceivingly simple at the first glance,
neither its solution for a seemingly ordinary situation nor the
assignment of appropriate exchange integrals $J_{ij}$ for a specific
material are trivial in any way. It is obvious that the crystal
structure of a compound is the key to understand its magnetic
properties. On the other hand, an assignment of interaction
parameters solely based on structural considerations can be completely
misleading like in the case of
(VO)$_2$P$_2$O$_7$.\cite{AHC_VO2P2O7_chiT,AHC_VO2P2O7_INS_chain_plus} In recent years, even
careful investigations based on accurate experimental data, but within
a limited spectrum of methods, have suggested controversially
discussed magnetic models for several compounds. A prominent example
for this problem are the two closely related spin-1/2 $J_1$-$J_2$ chain
compounds Li$_2$CuO$_2$ and LiCu$_2$O$_2$, for which consensus about
their location in the magnetic phase diagram was established only
recently.\cite{FHC_Li2CuO2_INS_chiT, lorenz2009, FHC_LiCu2O2_chiT_CpT_NS_wm_paper,
FHC_LiCu2O2_chiT_CpT_NS_wm_comment, FHC_LiCu2O2_chiT_CpT_NS_wm_paper_reply,
FHC_LiCu2O2_NMR_DFT, FHC_LiCu2O2_INS}

Thus, to establish the appropriate magnetic model for a new,
complex material, the application of independent methods
seems of crucial importance. In particular, the search for
the relevant sector in the phase diagram can largely benefit
from a detailed microscopic analysis based on modern band
structure
theory\cite{FSL_Li2VOSiO4_DFT,FHC_LiCu2O2_NMR_DFT,CuClLaNb2O7_DFT,CuClLaNb2O7_str_lowT}
in combination with numerical methods to solve subsequently
the corresponding Heisenberg Hamiltonian, at least in an
approximate way.\cite{kapel_hayd_DFT,volb_DFT}

In particular, for green dioptase a magnetic model with
antiferromagnetic (AFM) nearest-neighbor (NN) coupling $J_c$ along the
spiral chains (see Fig.~\ref{struc}, middle) and AFM coupling $J_d$
within the structural Cu$_2$O$_6$ dimers was suggested\cite{HC_CuSiO3H2O_simul_Raman} on
empirical grounds and evaluated using quantum Monte-Carlo simulations
(QMC) to fit the experimental magnetic susceptibility. This study
places the compound in proximity to a quantum critical point due to a
competition between chain-like ordering along $c$ and magnetic dimer
formation caused by the AFM $J_d$. In contrast to
Ref.~\onlinecite{HC_CuSiO3H2O_simul_Raman}, the results of our microscopic study place
the compound in a different region of the phase diagram and assign the
strong quantum fluctuations and the related magnetic properties to the
small effective coordination number of the magnetic Cu$^{2+}$ sites.
\section{Methods}


Electronic structure calculations were performed using the full potential
non-orthogonal local-orbital minimum basis scheme
\texttt{fplo9.00-33}.\cite{FPLO} For the scalar
relativistic calculations within the local density
approximation (LDA) the exchange and correlation potential
of Perdew and Wang was chosen. A well-converged $k$-mesh of
8$\times$8$\times$8 points was used for LDA calculations.
\cite{PW} Wannier functions (WF) were calculated for the
antibonding Cu $3d_{x^2-y^2}$ states. Strong correlations
are treated in a mean-field way within the LSDA+$U$
approach.\cite{FPLO_LSDU} For the double counting correction
(DCC) we applied the two limiting cases: the around-mean-field
(AMF) approach and the fully-localized limit
(FLL).\cite{LDA_U_AMF_FLL} The on-site Coulomb repulsion
$U_{3d}$ was varied within the physically reasonable ranges: $U_{3d}$=5.5--7.5~eV
for AMF and $U_{3d}$=6.5--9.5~eV for FLL. The intra-atomic Hund's coupling
$J_{3d}$ was fixed to 1~eV. To allow for various spin ordering arrangements,
the original hexagonal symmetry was reduced to the
spacegroup $P1$. For the LSDA+$U$ calculations, we used
$k$-meshes of 4$\times$4$\times$4 points. The calculations
were carefully checked for convergence. For the structural
input, we used the crystal structure from
Ref.~\onlinecite{HC_CuSiO3H2O_ENS}. 

Quantum Monte-Carlo (QMC) simulations were performed using
the programs \texttt{looper} and \texttt{dirloop\_sse}  of the
software package \texttt{ALPS}.\cite{ALPS} The magnetic
susceptibility was simulated for $N$=$10752$ sites clusters,
containing 256 coupled chains of 42 sites each. In the
temperature range $T/J_c=0.15$--4.50, we used 25 000 sweeps for
thermalization and 300 000 sweeps after thermalization.  The
resulting statistical errors ($<$0.1\%) are far below the
experimental inaccuracy. To evaluate the dependence of the
static structure factor on the cluster size, we performed a
series of simulations starting with a $N$=24 sites cluster
and consequently increasing it up to $N$=8232 sites.
Magnetization curves were simulated on $N$=1536 sites
clusters at $T$=0.025$J_c$ using 50 000 sweeps for
thermalization and 500 000 sweeps after thermalization.
Statistical errors did not exceed 0.5\%.

The experimental data were collected on a natural sample of
green dioptase. A green transparent
crystal was mechanically detached from the calcite matrix 
and used for magnetic measurements without alignment in the
magnetic field. The magnetic susceptibility was measured with a
Quantum Design MPMS SQUID in the temperature range 2--380~K
in applied fields up to 5~T.

\section{Results}
\subsection{Electronic structure and magnetic model}

\begin{figure}
\begin{center}\includegraphics[width=8cm,angle=0]{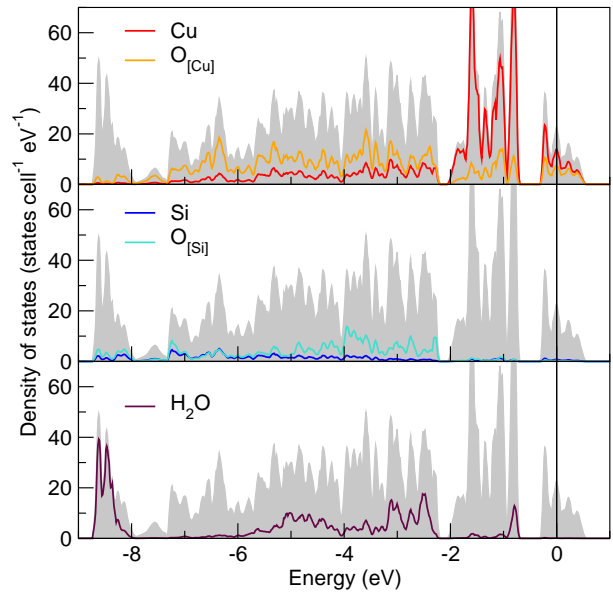}\end{center}
\caption{\label{dos}(Color online)
Total and atom-resolved LDA density of states for the green
dioptase.  The antibonding Cu--O $dp\sigma$ states form a
well separated band complex at the Fermi level.
}\end{figure}

The electronic structure of the green dioptase was calculated within
the LDA. The atom-resolved density of states (DOS)
is depicted in Fig.~\ref{dos}.  The width of the valence band
is about 10\,eV, similar to other cuprates.  States at the Fermi level
evidence a metallic solution in contrast to the green transparent
crystals indicating an insulating behavior. This well known
shortcoming of the LDA approach originates from the underestimation of
the strong Coulomb repulsion in the Cu $3d$ shell. The insulating
ground state can be restored by adding the missing part of correlation
(i) via mapping onto a Hubbard model or (ii) in a mean-field way by
LSDA+$U$ calculations. In this study, both approaches are used.

Despite the incorrect description of the ground state, LDA is known as
a reliable tool for the evaluation of relevant orbitals and
couplings. We start the analysis from the highest-lying states of the
valence band. The well-separated band complex at the Fermi level is
half filled and formed by antibonding Cu--O $dp\sigma$ states.  The
energy range between $-0.5$ and $-2$~eV is dominated by non-bonding O
and Cu states. At lower energies, around $-$2\,eV, states of
the SiO$_4$
tetrahedra and H$_2$O appear.  The sizable H$_2$O contribution
originates from the apical position of the crystal water in the
distorted CuO$_4$(H$_2$O)$_2$ octahedra, and thus rather short
Cu--O$_{\mathrm{H}_2\mathrm{O}}$ distances of 2.51 and 2.66~\AA.

\begin{figure}
\begin{center}\includegraphics[width=8cm,angle=0]{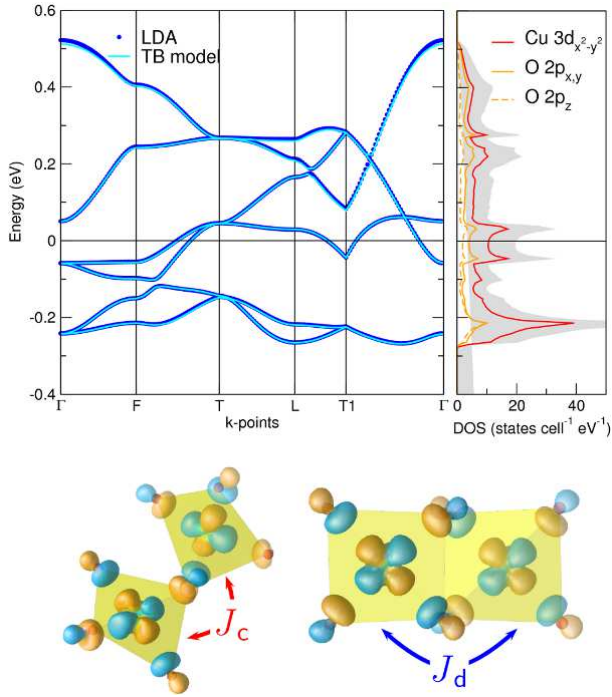}\end{center}
\caption{\label{band_wf}(Color online)
Top: Comparison of the antibonding $dp\sigma$ bands from the LDA
calculation and the tight-binding model (left) together with the
orbital-resolved density of states (right).  Bottom: Cu-centered
Wannier functions superimposed upon the leading superexchange
pathways. }\end{figure}

Typical for cuprates, the magnetic properties of the green dioptase
are ruled by the half-filled antibonding Cu--O $dp\sigma$ band complex
at the Fermi level. The width of this complex $W$ can be used as a
rough estimate for the leading couplings. Thus, $W$=0.8\,eV for the
green dioptase is comparable to related systems such as
Li$_2$ZrCuO$_4$ (buckled edge-shared chains,
$W$=0.5\,eV),\cite{FHC_Li2ZrCuO4_DFT_simul} Cu$_2$(PO$_3$)$_2$CH$_2$
(distorted dimers, $W$=1\,eV)
\cite{Cu2PO32CH2_DFT_NMR_chiT_CpT_MH_simul} or kapellasite
(kagome lattice of corner-shared plaquettes with a Cu--O--Cu
bond angle of about 107$^\circ$, $W$=0.9\,eV)
\cite{kapel_hayd_DFT}, but it is strongly reduced compared to
Sr$_2$CuO$_3$ (chains of corner-shared plaquettes,
$W$=2.5~eV)\cite{HC_Sr2CuO3_Ca2CuO3_DFT_RPA} or SrCuO$_2$ (zigzag
chains of edge-shared plaquettes,
$W$=2~eV)\cite{HC_SrCuO2_DFT_ARPES}. Based on such simplified
comparative analysis, we can conjecture the leading couplings in
dioptase to be of the order of 100~K.

The orbitals which are relevant for the magnetism can be evaluated by
a projection onto a set of local atomic orbitals. For each plaquette,
one of the Cu--O bonds and a direction perpendicular to the plaquette
are considered as local $x$ and $z$ coordinate axes, respectively.
This way, we find that Cu states in the $dp\sigma$ band complex have
nearly pure Cu~$3d_{x^2-y^2}$ character (see Fig.~\ref{band_wf},
right), although the plaquettes forming the structural Cu$_2$O$_6$
dimers are considerably distorted. On contrary, the orbital-resolved
DOS for the O states shows a mixture of 2$p_{x,y}$ and 2$p_z$
states. This mixing is caused by a non-coplanar arrangement of the
neighboring dimers, sharing a common O atom.
\footnote{To prove that, we plotted the orbital-resolved DOS
for individual atoms. The states corresponding to the O atoms bridging
two Cu atoms in a dimer show a clearly planar character. For the O
atoms linking two neighboring dimers, the contributions from O $2p_x$
and O $2p_z$ are comparable, because the O $2p_z$ states correspond to
the O $2p_x$ states of the neighbor-dimer, while the states from O
2$p_y$ are negligible.} Thus, although the O 2$p_z$ contributions are
unusually high and seemingly hint at sizable O $2p_\pi$ contributions,
the states around the Fermi energy are clearly dominated by Cu-O
$dp\sigma$ states. Since the number of bands forming the band complex
coincides with the number of plaquettes in the unit cell, magnetic
properties of the compound can be described by an effective one-band
TB model.

In case of the green dioptase, the evaluation of its
magnetic model from simple geometric considerations based on
the crystal structure only is difficult due to a complex 3D
coupling of the structural dimers. To develop a magnetic
model of the compound from microscopic grounds, the six
bands forming the complex at the Fermi level were mapped onto an
effective one-band TB model parameterized with transfer
integrals $t_{ij}$. The Wannier function (WF) technique yields an unambiguous
solution of this six-band problem. The resulting fit shows a
nearly perfect description of the LDA band structure
(Fig.~\ref{band_wf}).

Only two of the resulting transfer integrals $t_{ij}$ are relevant:
$t_c=126$\,meV (the subscript $c$ stands for ``chain''), running along
the spirals of dimers (in the $c$ direction), and $t_d=104$\,meV, the
intra-dimer coupling (compare Fig.~\ref{struc}, middle).  Other
hoppings (except for the hopping $t_{ic}$=$24$~meV between the
spirals) are smaller than 20~meV. To restore the insulating ground
state, we map the TB model onto a Hubbard model considering an
effective (one band) Coulomb repulsion $U_{\mathrm{eff}}=4$\,eV. For
the strongly correlated limit at half-filling, both well justified for
the green dioptase, the lowest lying (magnetic) excitations can be
efficiently described by a Heisenberg model. This way, the resulting
magnetic exchange can be derived using the second order perturbation
theory expression
$J^{\mathrm{AFM}}_{ij}$=$4t_{ij}^2/U_{\mathrm{eff}}$. Since the
original TB-model is a one-band model, only the antiferromagnetic
contribution to the total magnetic exchange is accounted for in this
approach. Thus, the resulting antiferromagnetic contributions for the
leading couplings are $J^{\mathrm{AFM}}_c$=184~K and
$J^{\mathrm{AFM}}_d$=125~K. Since exchange integrals
$J^{\mathrm{AFM}}_{ij}$ are proportional to $t^{2}_{ij}$, all further
exchanges are smaller than 7~K (less than 4\% of the leading exchange)
and can be neglected in first place.

Due to their close vicinity to 90$^\circ$
the intra-dimer Cu--O--Cu bond angle of 97.4$^{\circ}$ and inter-dimer
angle of 107.6$^{\circ}$ require a careful estimation of the
ferromagnetic contributions to the total exchange integrals, neglected
in the effective one-band TB model approach presented above. Thus, we
performed LSDA+$U$ calculations of magnetic supercells with various
collinear spin arrangements. The mapping of total energy differences
onto a classical Heisenberg model results in an AFM exchange along the
spirals chains $J_{c}$=110~K and a FM intra-dimer exchange
$J_{d}$=$-66$~K for a typical value of $U_{d}$=6.5~eV within the AMF
DCC scheme.

In agreement with the expectations according to the
Goodenough-Kanamori-Anderson rules,\cite{GKA_1,GKA_2,GKA_3} the LSDA+$U$
calculations evidence considerable FM contributions to both relevant
exchange integrals $J_c$ and $J_d$. In particular, the total value of
$J_c$ is strongly reduced compared to the estimate from the one-band
model ($J^{\mathrm{AFM}}_c$=184~K), yielding $J^{\mathrm{FM}}_c$=$-$74~K.
For $J_d$, the closer proximity of the Cu--O--Cu angle to
90$^{\circ}$ leads to an even larger FM contribution
$J^{\mathrm{FM}}_d$=$-$191~K which exceeds the AFM part
$J^{\mathrm{AFM}}_{d}$, resulting in a significantly FM total coupling
$J_d$ within the structural dimers.

\begin{figure}
\begin{center}\includegraphics[width=8cm]{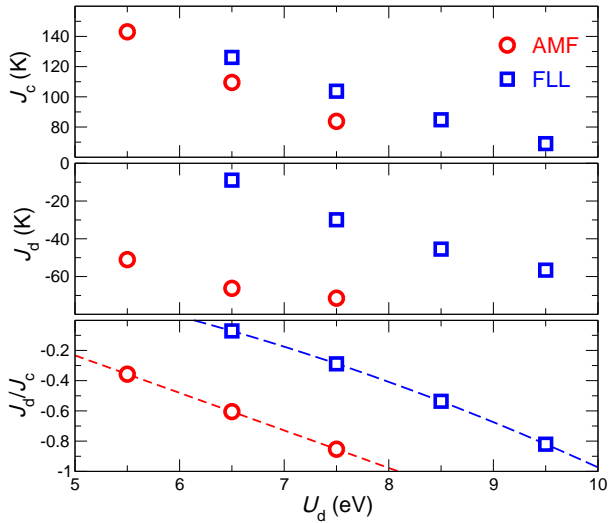}\end{center}
\caption{\label{lsdu}(Color online)
Results of total energy LSDA+$U$ calculations: the leading exchange
integrals ($J_c$ and $J_d$) and their ratio as a function of the
on-site Coulomb repulsion $U_{3d}$ for the around-mean field (AMF) and
the fully localized limit (FLL) double counting correction
schemes. }\end{figure}

Since the choice of DCC is non-trivial and can have a large impact on the
resulting exchange parameters\cite{HC_CdVO3_DFT,volb_DFT}, we compare
the AMF results to FLL calculations.  For the green dioptase, we find
that both DCC schemes yield similar results (Fig.~\ref{lsdu}). The
only apparent difference is related to the values of $U_{3d}$: for
FLL, about 2\,eV larger $U_{3d}$ values are required in order to
obtain the same exchange integrals as AMF. The FM nature of $J_d$
remains stable in the whole range of $U_{3d}$ and for the two DCC
schemes.

Although the qualitative microscopic model is well justified
by varying the $U_{3d}$ parameter in a rather wide range, the
strong dependence of the resulting exchange integrals on
$U_{3d}$ impedes an accurate estimation of the absolute size
and the ratio of the two couplings. In the next section, we
refine the values of the exchange integrals by alternative methods.
  
To summarize the microscopic analysis, we obtain a model with two
leading interactions: an AFM $J_c$ running along the spiral (in the
$c$ direction) and an FM $J_d$ inside structural Cu$_2$O$_6$
dimers. We should note that a related magnetic model was proposed in
Ref.~\onlinecite{HC_CuSiO3H2O_simul_Raman}.  It is based on the same
relevant exchange interactions, but implies an AFM intra-dimer
coupling $J_d$ in contrast to the FM nature of this coupling in our
model. Remarkably, a model very similar to ours has been used to
describe neutron scattering data for the dehydrated, black species of
dioptase Cu$_6$Si$_6$O$_{18}$. This issue will be discussed in
Sec.~\ref{disc}.

\subsection{Experimental results and model simulations}
\label{simulations}

\begin{figure}[tb]
\includegraphics[angle=270,width=8.6cm]{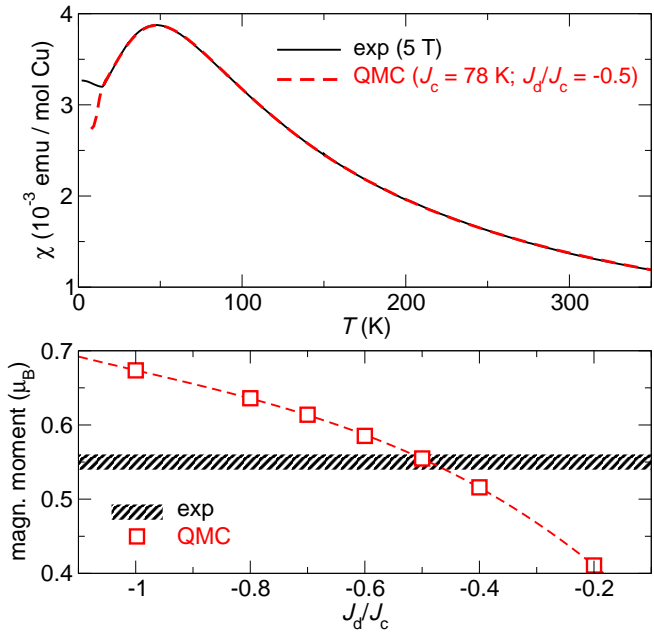}
\caption{\label{chiT_Mom}(Color online) Top: The quantum
Monte-Carlo (QMC) fit to the experimental magnetic
susceptibility (per mol Cu). Bottom: QMC results for the ordered magnetic
(sublattice) moment as a function of the $J_d/J_c$ ratio. The
experimental value from neutron diffraction\cite{HC_CuSiO3H2O_ENS} with
an error bar is depicted by a striped bar.}
\end{figure}

To challenge our model with respect to the experimental data, we
measured the magnetic susceptibility of the green dioptase
(Fig.~\ref{chiT_Mom}) and used neutron diffraction results from
previous studies.\cite{HC_CuSiO3H2O_ENS} Our susceptibility data are
in good agreement with the earlier
reports\cite{HC_CuSiO3H2O_ESR_chiT,HC_CuSiO3H2O_simul_Raman} and show
a broad maximum at 50~K along with the magnetic ordering anomaly,
evidenced by a kink at $T_N\simeq 15$~K.  We do not observe any
appreciable field dependence up to 5~T. Above 200~K, the
susceptibility can be fitted to the Curie-Weiss law, modified with an
additional temperature-independent contribution $\chi_0$, responsible
for core diamagnetism and van Vleck paramagnetism. The Curie-Weiss fit
leads to $\chi_0$=$-7.2\cdot10^{-5}$~emu/mol, the effective magnetic
moment $\mu_{\text{eff}}$=$1.99$~$\mu_B$ ($g$=2.30), and the Weiss
temperature $\theta$=43~K.  Note that the susceptibility below $T_N$
depends on the crystal orientation in the magnetic field and can only
be treated within an anisotropic model. In contrast, the magnetic
behavior in the paramagnetic regime (above $T_N$) is isotropic: the
susceptibility simply scales due to different $g$-values along
different crystal directions.\cite{HC_CuSe2O5_DFT_chiT_CpT_simul}
Since our microscopic study yields the exchange integrals of the
isotropic (Heisenberg) Hamiltonian, we restrict ourselves to the data
above $T_N$.

DFT calculations evidence two relevant interactions --- the AFM
intra-chain exchange $J_c$ and the FM intra-dimer (inter-chain)
exchange $J_d$. Below, we will compare this model to the experimental
results and refine the values of the leading exchange integrals.
Since the microscopic model is well justified qualitatively (relevant
couplings and their sign), its internal parameters $J_c$ and $J_d$ can
be refined by varying them in a reasonable range and subsequently
simulating the thermodynamical behavior for a given $J_d/J_c$ ratio.

A method for simulations should be certainly consistent with
the spin model. Since the two relevant couplings in dioptase
form a non-frustrated, formally three-dimensional spin
lattice, QMC simulations are natural and probably the only
feasible choice. Therefore, we perform QMC simulations for
the relevant parameter range $-1{\leq}J_d/J_c{\leq}-0.2$ of
the $J_c$-$J_d$ model.

Simulations of the Heisenberg Hamiltonian yield a reduced
magnetic susceptibility $\chi^{*}$ which is related to the
experimentally measured $\chi$ by the expression:

\begin{equation}
\chi(T)=
\frac{N_Ag^2{\mu}_{B}^2}{k_BJ_c}\cdot\chi^{*}\biggl(\frac{T}{k_BJ_c}\biggl)
+ \frac{C_{\text{imp}}}{T} + \chi_0,
\end{equation}

where $N_A$, $k_B$ and $\mu_B$ stand for the Avogadro
constant, the Boltzmann constant and the Bohr magneton,
respectively, $g$ is the Lande factor, $C_{\text{imp}}$ is the Curie
constant to account for possible impurity and defect contributions and
$\chi_0$ is a temperature independent term, similar to the Curie-Weiss fit.

Although a fit to $\chi(T)$ is commonly regarded as a sensitive probe
for internal parameters of a magnetic model,\footnote{We should note,
that a good fit to the experimental magnetic susceptibility is by no
means an evidence that the model itself is correct, since essentially
different sets of parameters for the same spin lattice and even for
different lattices can yield similar macroscopic magnetic behavior.}
we find that the ratio $J_d/J_c$ can be varied in a rather wide range
($-0.8$...$-0.4$) yielding a very good fit to the experimental data
for the paramagnetic phase~\footnote{For the ordered phase, anisotropy
plays a crucial role. Since the simulated model is isotropic, the
temperature range corresponding to the AFM ordered phase (below
$T_N<15$~K) was excluded from the fit.}. To improve our refinement of
$J_d/J_c$, we have to address the magnetic ordering temperature
$T_{N}$, which can be traced by a clear kink in the simulated
curves. The reference to $T_{N}$ yields $J_d/J_c$ close to $-0.5$. The
respective fit is shown in Fig.~\ref{chiT_Mom} (top). The resulting
$J_c$=78~K agrees well with the DFT estimates: 110~K for $U_{3d}$=6.5
eV within the AMF scheme ($J_d$/$J_c$=$-0.6$) and even better with
85~K yielded by $U_{3d}$=8.5 eV for the FLL scheme
($J_d/J_c$=$-0.55$). Moreover, $g=2.26$ and
$\chi_0$=$-6.9{\cdot}10^{-5}$ emu/mol are consistent with estimates
from the Curie-Weiss fit (2.30 and $-7.2{\cdot}10^{-5}$ emu/mol,
respectively).

For a further test of our model, we will address its ground state
properties. First, the propagation vector $\vec{q}$ of the AFM ordered
GS coincides with the experimentally observed
$\vec{q}$=$(0,0,\frac{2}{3}\pi)$\cite{HC_CuSiO3H2O_ENS} in the whole
range $-1{\leq}J_d/J_c{\leq}-0.2$. In this GS, the neighboring spins
along the spiral chains ($J_c$) align antiferromagnetically, while the
ordering within the edge-sharing dimers ($J_d$) is FM. This justifies
the validity of our microscopic model, but does not allow for a more
accurate refinement of the proposed value for the $J_d/J_c$ ratio. For
a further comparison, we use the sublattice magnetization ($m$) that
has been previously estimated in neutron diffraction experiments and
amounts to 0.55~$\mu_B$.\cite{HC_CuSiO3H2O_ENS}

Unfortunately, the theoretical estimation of $m$ is not
straightforward for two reasons.  First, the simulations do not yield
the magnetic moment in the ordered state directly.  Instead, it can be
estimated based on the static structure factor or on spin
correlations. Second, 3D coupling is crucial for the magnetic ordering
of the green dioptase, thus sizable finite size effects are expected
even for rather large clusters. To account for these effects, we use
the general procedure from
Ref.~\onlinecite{SL_QMC_finite_size_scaling} and estimate the magnetic
moment $m$ based on a finite-size scaling of the static structure
factor, taken for the propagation vector of the ordered structure. The
results of simulations for various $J_d/J_c$ ratios are shown in
Fig~\ref{chiT_Mom} (bottom). Remarkably, the theoretical $m$ for
$J_d/J_c$=$-0.5$ is in good agreement with the experimental value.

\begin{figure}[tb]
\includegraphics[angle=270,width=8.6cm]{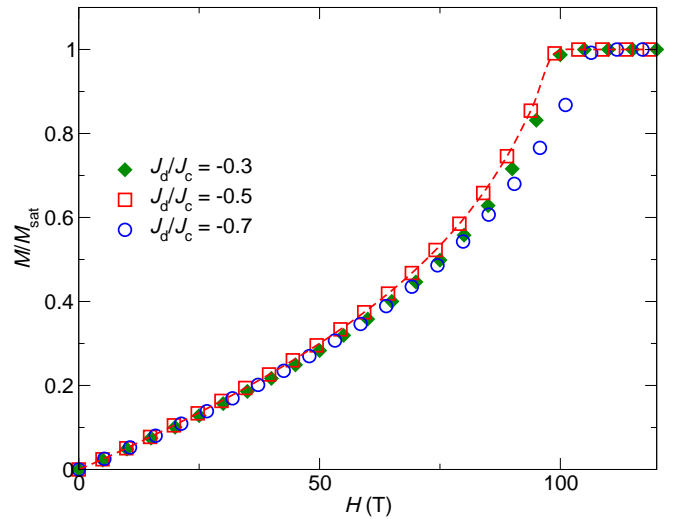}
\caption{\label{mh}(Color online) Simulated high field
magnetization curves for three different values of the $J_d/J_c$
ratio. The magnetic field is scaled by adopting the values of $J_c$ and
$g$ from the fits to $\chi(T)$.}
\end{figure}

Finally, we can introduce a magnetic field term to our Hamiltonian
and simulate the behaviour of magnetization $M$ as a function of
the reduced magnetic field $0{\leq}h^{*}{\leq}5J_d$. Such a simulation
could be an additional test for our model, since high-field
magnetization experiments were recently
announced.\cite{HC_CuSiO3H2O_ESR_chiT} Therefore, we simulate
$M(h^*)$ curves for $J_d/J_c$=$-0.5\pm0.2$ and scale them using the
expression

\begin{equation}
 M(h) = M\biggl(\frac{k_BJ_c}{g{\mu}_{B}}h^*\biggl),
\end{equation}

adopting the $J_c$ and $g$ values from the fits to $\chi(T)$. The
resulting curves shown in Fig.~\ref{mh} have similar shape and only
slightly different values of the saturation field.\footnote{For
non-frustrated spin lattice in dioptase, the value of the saturation
field depends on $J_c$ and $g$, but is not affected by the $J_d/J_c$
ratio in case of a FM $J_d$} Therefore, the experimental $M(h)$
dependence is unlikely to facilitate a further refinement of the model
parameters due to practical resolution limits at high magnetic fields.
In addition, the predicted value of the saturation field remains
challenging for present-day experimental facilities.

\section{\label{disc}Discussion}
In our model, the spin lattice of the green dioptase comprises AFM
couplings $J_c$ between the corner-sharing CuO$_4$ plaquettes and FM
couplings $J_d$ between the edge-sharing plaquettes
(Fig.~\ref{struc}). This situation is not surprising, because the
corner-sharing connection normally leads to $180^{\circ}$
superexchange, while the edge-sharing connection corresponds to the
Cu--O--Cu angle close to $90^{\circ}$. However, the crystal structure
of the green dioptase shows a tiny difference between the
superexchange pathways. The twisted configuration of the
corner-sharing plaquettes leads to the Cu--O--Cu angle of
$107.6^{\circ}$ for $J_c$ that is substantially larger than $97.4^{\circ}$
for $J_d$. The smaller angle for $J_d$ still fits to the general
trend, predicted by Goodenough-Kanamori-Anderson rules.\cite{GKA_1,GKA_2,
GKA_3} On the other hand, the green dioptase is very close
to the ``critical
regime'' of the Cu--O--Cu superexchange. Then, even a weak structural change
could lead to a strong modification of the exchange couplings, making an
empirical assignment of the parameter region difficult. For example, the
earlier theoretical analysis assumed AFM coupling for both $J_c$ and
$J_d$.\cite{HC_CuSiO3H2O_simul_Raman} Therefore, we used a
quantitative microscopic approach and demonstrated that this empirical
assumption is not consistent with the electronic structure of the
compound.

To examine whether small changes in the crystal structure may lead to
a modification of the microscopic model, the consideration of
structurally closely related compounds is a natural approach. The dehydration
transforms the green dioptase into the black dioptase
Cu$_6$Si$_6$O$_{18}$ that essentially keeps the 3D framework-type
crystal structure (Fig.~\ref{struc}) but lacks water molecules. The
Cu--O--Cu angles amount to $110.7^{\circ}$ and $97.3^{\circ}$ for
$J_c$ and $J_d$, respectively.\cite{breuer1989} Thus, based on basic
structural arguments, the signs of the two couplings should persist,
while the absolute values are likely increased. This prediction is in
line with the experimental data, indicating a large Weiss temperature
$\theta$ of 180~K\cite{wintenberger1993} of the black
dioptase compared to $\theta$=43~K in
the green dioptase (Sec.~\ref{simulations}). In addition, neutron
diffraction studies evidence similar magnetic structures for the black
and green dioptase.\cite{wintenberger1993} On the other hand, a very
recent study of black dioptase based on extended H\"uckel calculations
and a 1D fit to the magnetic susceptibility assigns the compound to
the model family of AFM chain compounds with very weak interchain
interactions.\cite{HC_bCuSiO3_chiT_CpT_DFT_str}

Further examples of the dioptase structure are given by the hydrated and
anhydrous Cu germanates \mbox{Cu$_6$Ge$_6$O$_{18}\cdot x$H$_2$O} with $x=0$
and~6. These compounds were previously considered as coupled frustrated spin
chains, because a sizable next-nearest-neighbor coupling along the spiral
chains was assumed.\cite{hase2003,hase2009} This assumption is rather empirical
and mainly motivated by the chemical similarity to the well-known spin-Peierls
compound CuGeO$_3$ with its frustrated spin chains of edge-sharing CuO$_4$
plaquettes.\cite{cugeo3} However, the pronounced difference in the crystal
structures strongly impedes a reliable transfer of the well established magnetic
model of the chain compound CuGeO$_3$ to the 
Ge-dioptase \mbox{Cu$_6$Ge$_6$O$_{18}$}.\cite{hase2003}  Based on
the results for the Si-dioptase, we would expect sizable AFM $J_c$, while $J_d$
is either FM or AFM. In the case that the FM and AFM contributions to $J_d$ are
close to cancel each other, the inter-chain coupling is effectively switched
off, and long-range couplings along the spiral chains could alter the physics.
The above empirical analysis gives no clear evidence for significant long-range
couplings, but further microscopic studies should challenge this conclusion.
Thus, since the minor structural changes in the dioptase family might be
crucial for changes in the leading magnetic couplings and the understanding of
their magnetic properties, a detailed comparative study is underway.\cite{future}

Taking the green dioptase as an example, we have derived the basic
features of the dioptase spin lattice. This spin lattice is
unfrustrated, hence we should preclude any references to the
frustrated spin chain model, at least for the green dioptase
Cu$_6$Si$_6$O$_{18}\cdot 6$H$_2$O. It is worth to mention that the
dioptase structure does \emph{not} give rise to the star lattice (decorated
honeycomb lattice), as it may seem on the first glance.\cite{richter}
Such confusion could arise from a specific projection of the crystal
structure, where the spiral chains look like flat frustrated triangles
(compare to the middle panel of Fig.~\ref{struc}).

After shortly outlining what the dioptase spin lattice is not, it is
more important to establish what it actually is: uniform AFM spin
chains aligned along the $c$ direction are arranged on the honeycomb
lattice, i.e., each chain is coupled to three neighboring chains, and
the system is geometrically 3D (Fig.~\ref{struc}). However, the
\emph{total} coordination number is as low as three: each atom has two
$J_c$ bonds and one $J_d$ bond only. Thus, the couplings in the $ab$
plane form a kind of a ``sparse'' honeycomb lattice. The reduction in
the coordination number has strong effect on the magnetic properties.

Experimental data for the green dioptase evidence strong quantum
fluctuations: the broad susceptibility maximum at
$T_{\max}^{\chi}/J_c\simeq 0.64$, the low N\'eel temperature
($T_N/J_c\simeq 0.2$), and the reduced sublattice magnetization
(0.55~$\mu_B$ compared to 1~$\mu_B$ for the classical spin-$\frac12$
systems). Strong quantum fluctuations are usually observed in
low-dimensional and/or frustrated spin systems. For example, the
archetypal two-dimensional spin model of the square lattice reveals
the susceptibility maximum at $T_{\max}^{\chi}/J\simeq 1.0$ and a
sublattice magnetization of
0.6~$\mu_B$.\cite{SL_QMC_finite_size_scaling} To reduce the ordering
temperature down to $T_N/J=0.2$, a very weak interlayer coupling
$J_{\perp}/J\sim 10^{-4}$ is required.\cite{siurakshina2000} Thus, the
quantum fluctuations in the dioptase spin lattice are even stronger
than in the square lattice, despite the 3D geometry.

Quantum fluctuations in a 3D spin system can arise from the magnetic
frustration (see Ref.~\onlinecite{nath2008} for an instructive
example). However, \emph{the dioptase spin lattice is neither
low-dimensional, nor frustrated}, hence its quantum behavior has a
different origin. We suggest that the long-range magnetic ordering in
dioptase is impeded by the low coordination number of the lattice,
because the low number of bonds reduces the exchange energy that
should stabilize the ordered ground state. The dioptase lattice can
thus be compared to low-dimensional spin systems with similar
coordination numbers. For example, the honeycomb lattice having three
bonds per site reveals the low sublattice magnetization of
0.54~$\mu_B$ and $T_{\max}^{\chi}/J\simeq 0.7$ (compare to 0.6~$\mu_B$
and 1.0 for the square lattice with four bonds per
site).\cite{loew2009} The apparent similarity between the dioptase and
the honeycomb lattice clearly shows that the coordination number is
the actual criterion of the ``low-dimensionality'', as long as the
magnitude of quantum fluctuations (the tendency towards the quantum
behavior) is considered. Although the conclusion is a natural
consequence of simple energy considerations, this is often
overlooked. While neither the dioptase crystal structure, nor its spin
model look low-dimensional, the essential physics is governed by
strong quantum fluctuations, typical for low-dimensional magnets. The
above considerations should stimulate further studies of
dioptase-structure materials and the respective spin model.

\section{Summary and Outlook}
Based on density functional calculations, quantum Monte-Carlo
simulations and magnetic measurements we have derived a new magnetic
model for the natural mineral green dioptase
Cu$_6$Si$_6$O$_{18}$$\cdot$6H$_2$O on a microscopic basis.  We have
shown that green dioptase can be well described by a quantum spin 1/2
Heisenberg model with essentially two relevant interactions: an NN AFM
intra-chain coupling $J_c \sim$ 78\,K within the spiral chains running
along the crystallographic $c$ direction, and a NN FM intra-dimer
(inter-chain) coupling $J_d \sim$ $-$37\,K within the structural
Cu$_2$O$_6$ dimers. The calculated temperature dependence of the
magnetic susceptibility, the magnetic ground state, the ordering
temperature and the sublattice magnetization for the suggested model
parameters are in very good agreement with the experimental data.
From our results we conclude that  the dioptase spin lattice is
neither low-dimensional nor frustrated, but exhibits large quantum
fluctuations due to a small effective coordination number of its magnetic
sites despite the three-dimensional lattice geometry.

Our approach demonstrates the great potential of the combination of
modern band structure methods and numerical simulations with magnetic
measurements for a reliable modeling of the magnetic properties for
complex compounds. An empirically based assignment of interaction
parameters for structurally complex systems can be easily misleading
and restrict studies to inappropriate regions of the magnetic phase
diagram.\cite{HC_CuSiO3H2O_simul_Raman} Since minor structural changes
may cause drastic changes in the leading magnetic couplings,
especially for couplings via bonds close to 90$^\circ$ relevant in the
dioptase family, a detailed comparative study for the hydrous and
anhydrous Si- and Ge-dioptase compounds is in progress.\cite{future}

\acknowledgements
We are grateful to Deepa Kasinathan for enlightening
discussions, Marcus Schmidt for providing us with the
samples of dioptase and Walter Schnelle for supporting 
the thermodynamical measurements and valuable comments.  A.
Ts. was supported by the postdoctoral fellowship of
the Alexander von Humboldt foundation.


%

\end{document}